 \documentstyle[proceedings,numreferences]{crckapb}

\newcommand{\ket}[2]{\vert #1 #2 \rangle}

\newcommand{\vet}[2]{\vert #1 #2)}

\newcommand{\tvet}[3]{\vert #1#2#3)}

\newcommand{\qvet}[4]{\vert #1#2#3#4)}

\newcommand{\qpn}[1]{[[#1]]_{qp}}
\newcommand{\srn}[1]{[[#1]]_{sr}}
\newcommand{\qqn}[1]{[[#1]]_{qq^{-1}}}

\newcommand{\Qn}[1]{[#1]_{Q}}
\newcommand{\Sn}[1]{[#1]_{S}}

\newcommand{\qn}[1]{[#1]_{q}}
\newcommand{\ti}[1]{\tilde #1}

\newcommand{\ud}{U_{qp}({\rm u}_2)}
\newcommand{\uduu}{U_{sr}({\rm u}_2) \supset {\rm u}_1}
\newcommand{\sud}{U_{q}({\rm su}_2)}
\newcommand{\sudQ}{U_{Q}({\rm su}_2)}
\newcommand{\suuu}{U_{q}({\rm su}_{1,1})}
\newcommand{\udm}{\frac{1}{2}}

\begin{opening}
\title{ON THE USE OF QUANTUM ALGEBRAS IN \protect \\
       ROTATION-VIBRATION SPECTROSCOPY}

\author{R. BARBIER}
\author{M. KIBLER }
\institute{Institut de Physique Nucl\'eaire de Lyon \\
 IN2P3-CNRS et Universit\'e Claude Bernard \\
           43 Boulevard du 11 Novembre 1918 \\
           F-69622 Villeurbanne Cedex, France}
\end{opening}

\runningtitle{QUANTUM ALGEBRAS AND SPECTROSCOPY}

\begin{document}
\begin{abstract}
A two-parameter deformation of the Lie
algebra u$_2$ is used, in conjunction
with the rotor system and the oscillator
system, to generate a model for rotation-vibration
spectroscopy of molecules and nuclei.
\end{abstract}
\section{Introduction}
Quantum algebras and  quantum groups \cite{Jimbo}
are nowadays widely used in physics. It is a point of fact
that most of the physical applications of quantum
algebras are based on one-parameter deformations.
Nevertheless, during the last years, multiparameter
(mainly two-parameter) quantum algebras have been
constructed [2-13].

In nuclear and molecular physics, one-parameter
quantum algebras have been used for describing
rotation spectra of nuclei [14-16]
 and rotation-vibration spectra of diatomic molecules [17-22].
In more details, we can distinguish three kinds of models. First,
 models describing only
rotation spectra in terms of the $\sud$ quantum algebra [14,~15,~17]
or the $q$-Poincar\'e algebra \cite{CGST}.
Second,  models describing only  vibration spectra
of diatomic molecules
and based on the $q$-deformation of the
${\rm u}_2 \supset {\rm o}_2$ chain \cite{BRF}, or on the
quantum algebra $\suuu$ \cite{BAR} or on the $q$-deformation
of the Heisenberg algebra $h_4$ \cite{CGY}.
Third, models describing rotation-vibration spectra
of diatomic molecules
in terms of a $q(j)$-deformation of the Heisenberg algebra
$h_4$  \cite{CY} or a $q$-deformation of the
${\rm u}_4 \supset {\rm so}_4 \supset {\rm so}_3$ chain
\cite{Pan}. Recently, a two-parameter
Hopf algebra \cite{Kibler1} has been applied to the
description of rotational spectra of superdeformed nuclei [23-25].

The work presented in this communication constitutes a first
step towards a model for describing simultaneously
rotation and vibration spectra of molecules and nuclei.
This model (to be developed in section 3) relies on the use
of two copies of the two-parameter quantum algebra $\ud$
(described in section 2). The approach is entirely of a
phenomenological nature although it is implicitly based on the
using of a non-rigid rotor system and an anharmononic oscillator
system.

\section{The Quantum Algebra $\ud$}
We begin with a brief description of the  quantum algebra $\ud$
restricted to those aspects necessary for the model
to be investigated here.
Two-parameter deformations of ${\rm su}_2$ and ${\rm u}_2$
have been considered by several authors. We
follow here the presentation of Refs.~[10,~12,~13].
According to Ref.~\cite{Kibler1}, the  quantum algebra $\ud$
is introduced by the following commutation relations
   \begin{equation}
        [J_0  ,J_\alpha] =  0                           \qquad
        [J_{3},J_{\pm} ] = \pm J_{\pm}                  \qquad
        [J_{+},J_{-  } ] = (qp)^{J_0-J_3} \> \qpn{2J_3}
   \label{eq:commutateur}
   \end{equation}
between four operators $J_\alpha$ ($\alpha = 0,3,+,-$).
In this work, we use  the notations
 \begin{equation}
\qpn{X}: = \frac{q^X - p^X}{q-p}      \qquad \quad
 \qn{X}: = \qqn{X} = \frac{q^X - q^{-X}}{q-q^{-1}}
 \label{eq:qp-qnombre}
 \end{equation}
 for the $qp$- and the $q$-deformations of a real number
 or an operator $X$.

The generators $J_\alpha$ verify the co-product rules
 \begin{eqnarray}
\Delta(J_0) &=& J_0  \otimes  1  +  1  \otimes  J_0
\label{eq:coproduit1} \\
\Delta(J_3) &=& J_3  \otimes  1  +  1  \otimes  J_3
\label{eq:coproduit2}  \\
\Delta(J_{+}) &=& J_{+}  \otimes
 (qp)^{J_0}  (qp^{-1})^{+J_3} + 1  \otimes  J_{+}
\label{eq:coproduit3}  \\
\Delta(J_{-}) &=& J_{-}  \otimes  1 +
(qp)^{J_0}  (qp^{-1})^{-J_3}  \otimes  J_{-}
 \label{eq:coproduit4}
 \end{eqnarray}
Note that (\ref{eq:coproduit3}) and (\ref{eq:coproduit4})
depends on the two independent parameters $q$ and $p$.
The universal ${\cal{R}}$-matrix
associated to the co-product $\Delta$ reads
\begin{equation}
{\cal{R}}_{qp} = \pmatrix{
p&0&0&0\cr
0&1&0&0\cr
0&p-q&pq&0\cr
0&0&0&p\cr
}
\label{eq:rmatrix1}
\end{equation}
in the representation
$\frac{1}{2} \otimes
 \frac{1}{2}$.

The operator \cite{Kibler1}
  \begin{equation}
C_2(U_{ {qp}}({\rm u}_2))\  =  \  {{1}\over{2}}(J_+J_- + J_-J_+)\
+\
{{1}\over{2}} \> \qpn{2} \> (qp)^{J_0-J_3} \>
\left( \qpn{J_3} \right)^2
 \label{eq:casimir1}
 \end{equation}
is invariant under the generators of the quantum algebra
$ U_{ {qp}}({\rm u}_2) $. It is
  characterized by  the judicious introduction of the element
$J_0$ which is itself a linear invariant of $ \ud $.

An irreducible representation of $\ud$,
in which the operators
$J_0$ and $C_2(U_{ {qp}}({\rm u}_2))$ are simultaneously diagonal
(in the case where neither $q$ nor $p$ are roots of unity),
is given by a Young pattern $[\varphi_1; \varphi_2]$ with
       $\varphi_1 - \varphi_2 = 2j  $ and
       $\varphi_1 + \varphi_2 = 2j_0 $ where
$2j_0$  and  $2j$  are  non-negative integers.
We introduce the vectors $\; \tvet{j_0}{j}{m}\; $
(with $m = - j, -j+1,\cdots,+j)$
as the basis vectors for the representation
$[j_0 + j \ ; \ j_0 - j ]$.
The vector $\; \tvet{j_0}{j}{m}\; $
is connected to the highest weight vector
                                  $\; \tvet{j_0}{j}{j}\; $
by the relation \cite{KAS}
 \begin{equation}
\tvet{j_0}{j}{m} \  = \  (qp)^{-\frac{1}{2}(j_0 - j)(j - m)} \;
\sqrt{
      \frac{ \qpn{j+m}!             }
           { \qpn{j-m}! \ \qpn{2j}! }
     } \;
(J_-)^{j-m} \ \tvet{j_0}{j}{j}
  \label{eq:base1}
  \end{equation}
Note that the unusual introduction of the
central element $J_0$ in (\ref{eq:base1})
comes from the commutation relations (\ref{eq:commutateur}) and
will play a crucial role in what follows.
By using (\ref{eq:base1}), we obtain the following action of the
generators  $J_\alpha$  on the state vector $\; \tvet{j_0}{j}{m} \;$
 \begin{eqnarray}
J_0    \ \tvet{j_0}{j}{m}  & = & j_0 \   \tvet{j_0}{j}{m}
 \nonumber \\
J_3    \ \tvet{j_0}{j}{m}  & = & m   \   \tvet{j_0}{j}{m}
 \nonumber \\
J_{\pm}\ \tvet{j_0}{j}{m}  & = &  (qp)^{\frac{1}{2}(j_0 - j)}
                             \sqrt{ \qpn{j \mp m} \qpn{j \pm m+1} }
                               \  \tvet{j_0}{j}{m \pm 1}
 \label{eq:actions}
 \end{eqnarray}
Note that Eq.~(\ref{eq:actions}) gives back the Jimbo \cite{Jimbo}
representation of $ \sud $ when $ p = q^{-1} $.

With the help of (\ref{eq:actions}), the eigenvalues
of (\ref{eq:casimir1}) in the irreducible representation
$[j_0 + j \ ; \ j_0 - j ]$ become
 \begin{equation}
 E(j_0,j)_{qp}  \; = \; (qp)^{j_0-j} \;  \qpn{j} \; \qpn{j+1}
 \label{eq:eigen1}
 \end{equation}
A factorisation of the
 eigenvalues (\ref{eq:eigen1})
 into a $j_0$-dependent and $j$-dependent part
is given by
 \begin{equation}
 E(j_0,j)_{qp}  \; = \; P^{2j_0-1} \; \Qn{j} \; \Qn{j+1}
 \label{eq:eigen1b}
 \end{equation}
where we use the notation (\ref{eq:qp-qnombre}) with
the parameters
 \begin{equation}
Q = (qp^{-1})^{1\over 2}  \qquad
P = (qp)     ^{1\over 2}
 \label{eq:paraQP}
 \end{equation}
as an alternative to $q$ and $p$.

The calculation of the Clebsch-Gordan coefficients
(CGc's) corresponding to the algebra homomorphism
$\Delta$~:~$ \ud \longrightarrow \ud \otimes \ud $
defined by
Eqs.~(\ref{eq:coproduit1}-\ref{eq:coproduit4})
is in progress. These coupling coefficients
depend on the two parameters $q$ and $p$.
It is to be emphasized that it seems not possible
to express them in terms of only
one of the two parameters $Q$ or $P$ given by
Eq.~(\ref{eq:paraQP}). In this respect, it is
perhaps interesting to note that, should we have
chosen, instead of $ \Delta $, the co-product
$\Delta_{qp}$ defined by
 \begin{eqnarray}
\Delta_{qp}(J_0) & = & J_0  \otimes  1  +  1  \otimes  J_0
\nonumber  \\
\Delta_{qp}(J_3) & = & J_3  \otimes  1  +  1  \otimes  J_3
\nonumber  \\
\Delta_{qp}(J_{\pm}) & = & J_{\pm}  \otimes
 {(qp)}^{{{1}\over{2}}J_0}  {(qp^{-1})}^{+{{1}\over{2}}J_3} +
 {(qp)}^{{{1}\over{2}}J_0}  {(qp^{-1})}^{-{{1}\over{2}}J_3}
                                    \otimes  J_{\pm}
 \label{eq:coproduit3b}
 \end{eqnarray}
we would have obtained CGc's depending only on the parameter
$Q$ (see Appendix).

All what preceds is of pivotal importance for application
to rotational spectroscopy. We now continue with
some aspects of $\ud$ useful in vibrational
spectroscopy.

The quantum algebra $\ud$ can be constructed from two
pairs, say $\{ {a}_+^+ ,{a}_+ \}$ and
           $\{ {a}_-^+ ,{a}_- \}$, of $qp$-deformed
(creation and annihilation) boson operators.
 The action of these $qp$-bosons
on a non-deformed two-particle Fock space
$ \{ \ket{n_+}{n_-} : n_+ \in {\bf N} , \
                      n_- \in {\bf N}  \} $
may be chosen to be controlled by
\begin{eqnarray}
{a}^+_+\;  \ket{n_+}   {n_-} & = &
              \sqrt{ \qpn{n_+ + \frac{1}{2} + \frac{1}{2}}} \;
              \ket{n_+ +1\ }{n_-}    \nonumber \\
{a}_+  \;  \ket{n_+}   {n_-} & = &
              \sqrt{ \qpn{n_+ + \frac{1}{2} - \frac{1}{2}}} \;
              \ket{n_+ -1\ }{n_-}    \nonumber \\
{a}^+_-\;  \ket{n_+}   {n_-} & = &
              \sqrt{ \qpn{n_- + \frac{1}{2} + \frac{1}{2}}} \;
              \ket{n_+}{\ n_- +1}    \nonumber \\
{a}_-  \;  \ket{n_+}   {n_-} & = &
              \sqrt{ \qpn{n_- + \frac{1}{2} - \frac{1}{2}}} \;
              \ket{n_+}{\ n_- -1}
\label{eq:qpbo}
\end{eqnarray}
The vectors $\ket{n_+}{n_-}$ are generated from
  \begin{equation}
 \ket{n_+}{n_-}  =
\frac{1}{\sqrt{\qpn{n_+}! \qpn{n_-}!}}  ({a}_+^+)^{n_+}
({a}_-^+)^{n_-} \; \ket{0}{0}
 \label{eq:base2}
  \end{equation}
The two pairs  $\{ {a}_+^+ ,{a}_+ \}$ and
                                        $\{ {a}_-^+ ,{a}_- \}$
of $qp$-bosons commute and satisfy [8,~26]
\begin{equation}
 {a}  _{\pm} {a}^+_{\pm}  =  \qpn{N_{\pm}+1}  \qquad \quad
 {a}^+_{\pm} {a}  _{\pm}  =  \qpn{N_{\pm}  }
\label{eq:commu}
\end{equation}
where $N_{+}$ and $N_{-}$ are the usual number operators with
\begin{equation}
N_{\pm} \ \ket{n_+}{n_-} =
n_{\pm} \ \ket{n_+}{n_-}
\label{eq:openomb}
\end{equation}
Of course, the $qp$-bosons
${a}^+_{\pm}$ and ${a}  _{\pm}$
reduce to ordinary bosons in the limiting situation
where $p=q^{-1} \rightarrow 1 $.

The passage from the (harmonic oscillator) state vectors
$\ket{n_+}{n_-}$ to angular momentum state vectors $\vet{j}{m}$
is achieved through the relations
\begin{equation}
 j := \frac{1}{2}(n_+ + n_-) \quad \qquad
 m := \frac{1}{2}(n_+ - n_-)
\label{eq:sch1}
\end{equation}
and
\begin{equation}
 \vet{j}{m}\  \equiv \ \ket{j+m\ }{j-m} \  = \  \ket{n_+}{n_-}
\label{eq:sch2}
\end{equation}
Equations (\ref{eq:qpbo}) may thus be rewritten as
 \begin{eqnarray}
{a}_{\pm}^+ \; \vet{j}{m} & = &
\sqrt{ \qpn{j{\pm}m+\frac{1}{2}+\frac{1}{2}}}\; \;
\vet{j+\frac{1}{2}\ }{m{\pm} \frac{1}{2}}    \nonumber \\
{a}_{\pm}   \; \vet{j}{m} & = &
\sqrt{ \qpn{j{\pm}m+\frac{1}{2}-\frac{1}{2}}}\; \;
\vet{j-\frac{1}{2}\ }{m{\mp} \frac{1}{2}}
\label{eq:qpbo2}
\end{eqnarray}
so that the $qp$-bosons behave as ladder operators for the
quantum numbers $j$ and $m$ (with $\vert m \vert \le j$).

Contact with the quantum algebra $\ud$ may now be
established. Let us define the four operators
$\ti{J}_\alpha$ ($\alpha = 0,3,+,-$)
as
\begin{equation}
\ti{J}_0   :=  {{1}\over{2}}(N_+ + N_-)  \quad
\ti{J}_3   :=  {{1}\over{2}}(N_+ - N_-)  \quad
\ti{J}_+   :=   {a}_+^+ {a}_-            \quad
\ti{J}_-   :=   {a}_-^+ {a}_+
\label{eq:gener1}
\end{equation}
A simple calculation shows that the operators $\ti{J}_\alpha$
satisfy the same  commutation relations as the operators
${J}_\alpha$ [see Eq.~(\ref{eq:commutateur})].
Consequently, Eq.~(\ref{eq:gener1}) provides us with a boson realization
of $\ud $. We may ask what kind of representation we thus obtain~?
Indeed, repeated application of Eqs.~(\ref{eq:qpbo2})
yields
 \begin{eqnarray}
\ti{J}_0     \ \vet{j}{m}  & = & j \   \vet{j}{m}
 \nonumber \\
\ti{J}_3    \ \vet{j}{m}   & = & m \   \vet{j}{m}
\nonumber  \\
\ti{J}_{\pm} \ \vet{j}{m}  & = &
\sqrt{ \qpn{j \mp m} \qpn{j \pm m+1} }
                                   \   \vet{j}{m \pm 1}
 \label{eq:actions3}
 \end{eqnarray}
so that the boson realization (\ref{eq:gener1}) concerns
the irreducible representation $[2j \ ; \ 0 ]$
for which $j_0 = j$ [cf.~Eq.~(\ref{eq:actions})].

At this stage, it is useful to mention the range of
variation of the parameters $q$ and $p$ of interest
for practical applications.
As a matter of fact, from Hermitean conjugation
requirements, the values of
the parameters
$q$ and $p$ must be restricted to some domains that can be classified
as follows: (i) $q \in {\bf R}$ and
                 $p \in {\bf R}$, (ii)
                 $q \in {\bf C}$ and
                 $p \in {\bf C}$ with the constraint $p=q^{\ast}$, and (iii)
                 $q = p^{-1} = {\rm e}^{{\rm i} \beta}$ with
$0 \le \beta < 2 \pi$.

\section{A Model for Rotation-Vibration Spectroscopy}
We are now in position to present a model
for describing rotation-vibration spectra
of diatomic molecules  and  nuclei.  This model
relies on two dynamical systems, viz., the non-rigid
rotor system with the dynamical symmetry $ \ud $
and the anharmonic oscillator system with the dynamical
symmetry $ \uduu $.
One of the
cornerstones of this model is the
Hamiltonian
 \begin{equation}
H   =  E_{\rm rovib} \ C_2( \ud ) \ + \ E_{\rm vib} \ C( \uduu ) \ + \ E_0
 \label{eq:casimir3}
 \end{equation}
where $ E_{\rm rovib} $ and $ E_{\rm vib} $ are two constants and
$ E_0  $ the zero energy term.
The first term $  C_2( \ud ) $ is the invariant (\ref{eq:casimir1})
of the quantum algebra $ \ud $ and thus is appropriate
for the rotational part.
We shall see how to make this term vibration-dependent
too by introducing a convenient coupling  with
the second term $ C( \uduu ) $.
The latter term describes the vibrational part.
We take it in the form
\begin{equation}
C( \uduu ) \ = \ \left( \srn{2{\ti{J}_0}} \right)^2 -
                (sr)^{2(\ti{J}_0-\ti{J}_3)}
                \left( \srn{2\ti{J}_3}  \right)^2
\label{eq:casimirvib}
  \end{equation}
where $ \ti{J}_0 $ and $ \ti{J}_3 $ are defined by
(\ref{eq:gener1}) and refer to a second quantum algebra
$ U_{sr}({\rm u}_2) $. The operator $ C( \uduu ) $
is invariant under the generators  $ \ti{J}_0 $ and $ \ti{J}_3 $
of the chain $ \uduu $. The Hamiltonian  $ H $ clearly
exhibits the  $ \ud \times (\uduu) $ dynamical symmetry.

The passage from the $sr$-boson state vectors $\ket{n_+}{n_-}$
[see Eq.~(\ref{eq:base2})] to the vibration state vectors
$\vet{n}{v}$ is accomplished by means of
 \begin{equation}
\vet{n}{v}\  \equiv \ \ket{n-v,}{v}
  \label{eq:vibstate}
  \end{equation}
In other words, we put
\begin{equation}
 n    = n_+ + n_-  \qquad
 n-2v = n_+ - n_-
 \label{eq:numstate}
  \end{equation}
The total number $n$ of $sr$-bosons is connected
to the maximal vibration quantum number  $v_{\rm max}$
by  $v_{\rm max} = n/2$ or $(n-1)/2$ depending wheather as
$n$ is even or odd. On the other hand, the rotational
state vectors are chosen to be $ \tvet{j_0}{j}{m} $
[see Eq.~(\ref{eq:base1})]. The coupling between
the rotational part and the vibrational
part is achieved by assuming that $ j_0 = v $.
This leads to the rotation-vibration
state vectors
\begin{equation}
\qvet{n}{v}{j}{m}  = \tvet{v}{j}{m} \otimes \vet{n}{v}
 \label{eq:rotvibstate}
  \end{equation}
where $v$ and $j$ are the vibrational
and rotational quantum numbers, respectively.
The diagonalisation of $H$ on the subspace spanned
by the vectors $\qvet{n}{v}{j}{m}$
leads to the eigenvalues
 \begin{equation}
E(n,v,j) = E_{\rm rovib} \> (qp)^{v-j} \> \qpn{j } \> \qpn{j+1   }
         + E_{\rm   vib}               \> \srn{2v} \> \srn{2(n-v)}
         + E_0
 \label{eq:eigen3}
 \end{equation}
Alternatively, Eq.~(\ref{eq:eigen3}) can be rewritten as
 \begin{equation}
E(n,v,j) = E_{\rm rovib} \> P^{2v-1  } \> \Qn{j } \> \Qn{j+1   }
         + E_{\rm vib  } \> R^{2(n-1)} \> \Sn{2v} \> \Sn{2(n-v)}
         + E_0
 \label{eq:eigen3b}
 \end{equation}
where we have introduced the parameters
 \begin{equation}
S = (sr^{-1})^{1\over 2} \qquad
R = (sr)     ^{1\over 2}
 \label{eq:paraRS}
 \end{equation}
to be compared with the parameters $Q$ and $P$ of Eq.~(\ref{eq:paraQP}).

The model described by Eq.~(\ref{eq:eigen3}) depends
on four quantum algebra parameters (namely, $q$, $p$, $s$ and $r$).
In the most general situation,
we thus have eight real parameters.
This number may be reduced following
the discussion at the end of section 2.
Here, we shall limit ourselves to the case
(ii) of the latter discussion for the pairs
($q$,$p$) and ($s$,$r$).
This yields four real parameters.
Furthermore, the consideration of limiting processes leads us to take
\begin{equation}
q = p^{\ast}=   {\rm e}^{\beta \cos \gamma} \;
                    {\rm e}^{+{\rm i} \beta \sin \gamma} \qquad
s = r^{\ast}=   {\rm e}^{\tau  \cos \gamma} \;
                    {\rm e}^{+{\rm i} \tau  \sin \gamma}
\label {eq:qprs}
\end{equation}
so that we end up with three real parameters $ \beta $, $ \gamma $ and $ \tau
$.
Then, Eq.~(\ref{eq:eigen3}) becomes
 \begin {eqnarray}
\lefteqn{ E(n,v,j)  =  E_{\rm rovib}
   \ {\rm e}^{(2v-1) \beta \cos \gamma} \;
      { {\sin (j \beta \sin \gamma) \; \sin [(j+1)
                                 \beta \sin \gamma]} \over
                                {\sin^2 (\beta \sin \gamma)} } }
 \nonumber \\
                              &   &
+ E_{\rm vib} \ {\rm e}^{(2n-2) \tau \cos \gamma} \;
      { {\sin (2v \tau \sin \gamma) \; \sin [2(n-v)
                                    \tau \sin \gamma]} \over
                                {\sin^2 (\tau  \sin \gamma)} } + E_0
 \label{eq:eigen4}
 \end{eqnarray}
Equation (\ref{eq:eigen4}) can be developed as
   \begin{equation}
E(n,v,j) \; = \; \sum_{l,k} \; Y_{lk} \; (v+\udm)^l \; [j(j+1)]^k
   \label{eq:dunham}
   \end{equation}
which resembles the Dunham \cite{Dunham} expansion
obtained from the Morse oscillator system.
The coefficients $Y_{lk}$ in Eq.~(\ref{eq:dunham})
depends on the parameters $\beta$, $\gamma$ and $\tau$
(and also $n$ which can be employed to
characterize the anharmonicity constant of
the oscillator).
They shall be  reported elswhere. It should be noted that
the limiting case $\gamma = \frac{\pi}{2}$ corresponds to
a model where the rotational and vibrational parts are
decoupled; in this case, the only non-vanhishing coefficients
$Y_{lk}$ in Eq.~(\ref{eq:dunham}) are of the type
$Y_{0k}$ and
$Y_{l0}$.

\section{Closing Remarks}
We have concentrated in this work on an $\ud \times (\uduu)$ model that unifies
and extends various models developed in recent years. The limiting
cases ($E_{\rm rovib} = 0$, $q = p^{-1} = s = r^{-1} = {\rm e}^{{\rm i} \tau}$)
and   ($E_{\rm rovib} = 0$, $q = p^{-1} = s = r^{-1} = 1$) correspond to the
$U_q({\rm u}_2) \supset {\rm o}_2$ model and
$    {\rm u}_2  \supset {\rm o}_2$ model worked out
in Ref.~\cite{BRF} and Ref.~\cite{RLD}, respectively,
for vibrational spectra of molecules, while the limiting case
($E_{\rm vib} = 0$, $q = p^{-1} = {\rm e}^{{\rm i} \beta}$)
corresponds to the $U_q({\rm su}_2)$ model introduced \cite{RRS}
for rotational
spectroscopy of nuclei. The particular case ($E_{\rm vib} = 0$, $v = j_0 = j$)
is nothing but the $\ud$ rotor model successfully applied to rotatinal
bands of superdformed nuclei \cite{BMK1,BMK2}.

A second important step towards the understanding of the dynamics
inherent to the $\ud \times (\uduu)$ model remains to be made.
This represents a difficult task involving some difference equations
and quantum inverse scattering methods. Another appealing project should be to
replace the Fock oscillator
states corresponding to Eqs.~(15-18) by the states
(which do not have classical limit) recently introduced by Rideau \cite{Rid}.
We hope to return on these matters in the future.

One of the authors (M.K.) would like to express his sincere gratitude to Guy
Rideau for his help, friendly advices and interesting discussions on various
occasions during the last ten years. Thanks are due to Guy Rideau and Pavel
Winternitz for useful comments at several stages of this work. Finally, M.K.
is indebted to Moshe Flato for friendly and constructive criticism.

\appendix
\section{Clebsch-Gordan Coefficients}
The CGc's
\begin{equation}
             (j_{01} j_{02} j_1 j_2  m_1 m_2 \vert j_0 j m )_{qp}
            \, \equiv \,
                                    (m_1 m_2 \vert       m )_{qp}
\label{eq:cgc1}
\end{equation}
corresponding to the co-product $\Delta_{qp}$ defined
through Eq.~(\ref{eq:coproduit3b})
can be seen to satisfy the following three-term recursion
relations \cite{KAS}
 \begin{eqnarray}
\lefteqn{ \sqrt{ [j \mp m ]_Q \ [j \pm  m + 1) ]_Q } \
          (m_1        m_2       \vert m \pm 1)_{qp}    } \nonumber \\
& &
 =  \ Q^{+ m_2} \ \sqrt{ [j_1 \pm m_1]_Q \ [j_1 \mp m_1 + 1]_Q } \
        (m_1 \mp 1, m_2       \vert m      )_{qp}        \nonumber \\
& & +  \ Q^{- m_1} \ \sqrt{ [j_2 \pm m_2]_Q \ [j_2 \mp m_2 + 1]_Q } \
        (m_1,       m_2 \mp 1 \vert m      )_{qp}
\label{eq:cgc2}
\end{eqnarray}
which are identical to the ones \cite{KCS} satisfied by the CGc's
$( j_1 j_2  m_1 m_2 \vert j m )_{Q} $ of the quantum
algebra $\sudQ$. Therefore, there exists a
proportionality constant between the $qp$-CGc's
and the $Q$-CGc's. For $q$ and $p$ real,
reality and normalization conditions can be used to
justify that the proportionality constant
is taken to be equal to 1. In fact, this may
be checked by direct calculation:
By adapting the method of projection operators
used for $\sud$ in Ref.~\cite{STK}, we can show that
\begin{equation}
(j_{01}j_{02}j_1j_2m_1m_2|j_0jm)_{qp} \; = \; \delta(j_0, j_{01} + j_{02}) \;
            (j_1j_2m_1m_2|   jm)_{Q }
\label{eq:cgc3}
\end{equation}
with
 \begin{eqnarray}
     \lefteqn{
(j_1j_2m_1m_2 |jm)_Q =  (-1)^{j_1 + j_2 - j}
       Q^{ - {1\over 2} (j_1+j_2-j)(j_1+j_2+j+1) + j_1m_2 - j_2m_1 }
     }
\nonumber \\
& &
 \bigg(
 {[j_1-j_2+j]! [j_1+j_2-j]! [j_1+j_2+j+1]! [j_2-m_2]! [j+m]!\over
  [j-j_1+j_2]! [j_2+m_2]! [j-m]! [j_1-m_1]! [j_1+m_1]!} \bigg) ^{1\over 2}
\nonumber \\
& &
 [2j+1]^{\udm}
 \sum_z \,
 {(-1)^{z} Q^{z(j_1+m_1)} [2j_2-z]![j_1+j_2-m-z]!\over
  [z]! [j_1+j_2-j-z]! [j_2-m_2-z]! [j_1+j_2+j+1-z]!}
\nonumber \\
\label{eq:cgc4}
 \end{eqnarray}
where we have used the abbreviation
$ \Qn{x} \equiv [x]$.

The results (\ref{eq:cgc3}) and (\ref{eq:cgc4})
can be also justified as follows. The universal
$R$-matrix associated to the co-product $\Delta_{qp}$
reads
\begin{equation}
{\cal{R}}_{pq} = \pmatrix{
p&0&0&0\cr
0&{(pq)}^{\udm}&0&0\cr
0&p-q&{(pq)}^{\udm}&0\cr
0&0&0&p\cr
}
\label{eq:rmatrix1b}
\end{equation}
which can be factorised in terms of $Q$- and $P$-depending parts.
Such a factorisation corresponds to the decomposition
\begin{equation}
U_{qp}({\rm u}_2) = {\rm u}_1 \otimes U_Q({\rm su}_2)
\label{eq:direct}
\end{equation}
where $\sudQ $ is spanned by
\begin{equation}
  A_0    :=  J_0   \qquad \quad
  A_3    :=  J_3   \qquad \quad
  A_\pm  := (qp)^{-{1\over 2}(J_0 - {1\over 2})} \, J_\pm
\label{eq:Aoper}
\end{equation}
with the following commutation relations
 \begin{equation}
[A_3,A_\pm   ] \ = \ \pm A_\pm   \qquad \quad
[A_+,A_-     ] \ = \ [2A_3]_Q
\label{eq:acomm}
\end{equation}
In terms of co-product, we have
 \begin{equation}
\Delta_{qp}(J_{\pm}) \> = \> P^{ \Delta_{Q}(A_0) - \frac{1}{2} }
\> \Delta_{Q}(A_{\pm})
\label{eq:cpfac}
\end{equation}
where the co-product $ \Delta_{Q} $ is given via
\begin{eqnarray}
\Delta_{Q}(A_0) & = & A_0 \otimes 1 + 1 \otimes A_0  \nonumber \\
\Delta_{Q}(A_3) & = & A_3 \otimes 1 + 1 \otimes A_3  \nonumber \\
\Delta_{Q}(A_{\pm}) & = &       A_{\pm} \otimes Q^{+A_3} +
                               Q^{-A_3} \otimes A_{\pm}
\label{eq:acopro}
\end{eqnarray}
which relations correspond to the CGc's $(j_1j_2m_1m_2 |jm)_Q$.


\begin{thebibliography}{99}

\bibitem{Jimbo}
  Kulish, P.P. and  Reshetikhin, N.Yu. (1981)
  {\it Zap. Sem. LOMI} {\bf 101}, 101
  [(1983) {\it J. Soviet. Math.} {\bf 23}, 2435];
  Sklyanin, E.K. (1982) {\it Funkt. Anal. Pril.} {\bf 16}, 27
  [(1982) {\it Funct. Anal. Appl.} {\bf 16}, 262];
  Drinfeld, V.G. (1985) {\it Soviet. Math. Dokl.} {\bf 32}, 254;
  (1986) in: {\it Proc. Int. Congr. Math.},
  Ed., A.M. Gleason (AMS, Providence, RI) pp. 798;
  Jimbo, M. (1985) {\it Lett. Math. Phys.} {\bf 10}, 63;
  (1986) {\it Commun. Math. Phys.} {\bf 102}, 537;
  Woronowiccz, S.L. (1987) {\it Publ. RIMS-Kyoto} {\bf 23}, 117;
                    (1987) {\it Commun. Math. Phys.} {\bf 111}, 613

  \bibitem{Sudbery}
  Sudbery, A. (1990) {\it J. Phys. A: Math. Gen.} {\bf 23}, L697

 \bibitem{DMMZ}
 Demidov, E.E., Manin, Yu.I., Mukhin, E.E. and Zhdanovich, D.V.
 (1990) {\it Prog. Theo. Phys. Suppl.} {\bf 102}, 203

 \bibitem{Resh}
 Reshetikhin, N. (1990) {\it Lett. Math. Phys.} {\bf 20}, 331

 \bibitem{FZ}
 Fairlie, D.B. and Zachos, C.K. (1991)
 {\it Phys. Lett. B} {\bf 256}, 43

 \bibitem{SWZ}
 Schirrmacher, A., Wess, J. and Zumino, B. (1991)
 {\it Z. Phys. C} {\bf 49}, 317

 \bibitem{Vokos}
 Vokos, S.T. (1991) {\it J. Math. Phys.} {\bf 32}, 2979

 \bibitem{CJ}
 Chakrabarti, R. and Jagannathan, R.
 (1991) {\it J. Phys. A: Math. Gen.} {\bf 24}, L711

  \bibitem{Dobrev}
  Dobrev, V.K. (1992) {\it J. Math. Phys.} {\bf 33}, 3419

  \bibitem{Kibler1}
  Kibler, M.R. (1993)
  in: {\it Symmetry and Structural Properties of Condensed Matter},
  Eds., W. Florek, D. Lipi\'nski and T. Lulek
  (World Scientific, Singapore) pp.~445

  \bibitem{CJ94}
  Chakrabarti, R. and Jagannathan, R. (1994)
  {\it J. Phys. A: Math. Gen.} {\bf 27}, 2023

  \bibitem{KAS}
  Kibler, M.R., Asherova, R.M. and Smirnov, Yu.F.
  (1994) preprint LYCEN 9439

  \bibitem{JV}
  Jagannathan, R. and Van der Jeugt, J. (1994) preprint hep-th/9411200

  \bibitem{Iwao}
  Iwao, S. (1990) {\it Prog. Theor. Phys.} {\bf 83}, 363

  \bibitem{RRS}
  Raychev, P.P., Roussev, R.P. and Smirnov, Yu.F.
  (1990) {\it J. Phys. G: Nucl. Phys.} {\bf 16}, L137

  \bibitem{CGST}
  Celeghini, E., Giachetti, R., Sorace, E. and Tarlini, M.
  (1992) {\it Phys. Lett. B} {\bf 280}, 180

  \bibitem{BRRS}
  Bonatsos, D., Raychev, P.P., Roussev, R.P. and Smirnov, Yu.F.
  (1990) {\it Chem. Phys. Lett.} {\bf 175}, 300

  \bibitem{BRF}
  Bonatsos, D., Raychev, P.P. and Faessler, A.
  (1991) {\it Chem. Phys. Lett.} {\bf 178}, 221

  \bibitem{BAR}
  Bonatsos, D., Argyres, E.N. and Raychev, P.P.
  (1991) {\it J. Phys. A: Math. Gen.} {\bf 24}, L403

  \bibitem{CGY}
  Chang, Z., Guo, H.Y. and Yan, H.
  (1991) {\it Phys. Lett. A} {\bf 156}, 192

  \bibitem{CY}
  Chang, Z. and Yan, H. (1991) {\it Phys. Lett. A} {\bf 158}, 242

  \bibitem{Pan}
  Pan, F.Z. (1993) {\it J. Phys. B: At. Mol. Opt. Phys.} {\bf 26}, L47

  \bibitem{BMK1}
  Barbier, R., Meyer, J. and Kibler, M.
  (1994) {\it J. Phys. G: Nucl. Phys.} {\bf 17}, L67

  \bibitem{Kibler2}
  Kibler, M. (1994)
  in: {\it Generalized Symmetries in Physics},
  Eds., H.-D. Doebner, V.K. Dobrev and A.G. Ushveridze
  (World Scientific, Singapore) pp.~55

  \bibitem{BMK2}
  Barbier, R., Meyer, J. and Kibler, M. (1994) preprint LYCEN 9437

  \bibitem{KK}
  Katriel, J. and Kibler, M. (1992)
  {\it J. Phys. A: Math. Gen.} {\bf 25}, 2683

  \bibitem{Dunham}
  Dunham, J.L. (1932) {\it Phys. Rev.} {\bf 41}, 721

 \bibitem{RLD}
 Van Roosmaleen, O.S., Levine, R.D. and Dieperink, A.E.
 (1983) {\it Chem. Phys. Lett.} {\bf 101}, 512

\bibitem{Rid}
Rideau, G. (1992) {\it Lett. Math. Phys.} {\bf 24}, 147

 \bibitem{KCS}
 Kibler, M., Campigotto, C. and Smirnov, Yu.F.
 (1994) in: {\it Proceedings of the International
 Workshop ``Symmetry Methods in Physics, in
 Memory of Professor Ya.A. Smorodinsky''},
 Eds., A.N. Sissakian, G.S. Pogosyan and S.I. Vinitsky,
 (JINR, Dubna, Russia) pp. 246

 \bibitem{STK}
 Smirnov, Yu.F., Tolsto\u \i, V.N. and Kharitonov, Yu.I.
 (1991) {\it Sov. J. Nucl. Phys.} {\bf 53}, 593

\end{thebibliography}
\end{document}